\documentclass{aa}  
\usepackage{natbib}
\usepackage{graphicx}
\usepackage[squaren,Gray]{SIunits}
\usepackage{color, colortbl}
\usepackage{txfonts}
\begin{document} 

\title{Prospects of detecting the polarimetric signature of the Earth-mass planet $\alpha$ Centauri B b with SPHERE / ZIMPOL}
\titlerunning{Detecting the polarimetric signature of $\alpha$ Cen B b with SPHERE / ZIMPOL}

   \author{
   	J. Milli   \inst{1,2}
          \and D. Mouillet   \inst{1}
          \and D. Mawet \inst{2}
	 \and  H. M. Schmid \inst{3}
  	  \and  A. Bazzon \inst{3}
  	  \and  J. H. Girard \inst{2}
  	  \and K. Dohlen \inst{4}
  	  \and R. Roelfsema \inst{3}
          }

   \institute{Institut de Planetologie et d'Astrophysique de Grenoble (IPAG), University Joseph Fourier,
              CNRS, BP 53, 
              38041 Grenoble, France\\
              \email{julien.milli@obs.ujf-grenoble.fr}
   		\and
  		European Southern Observatory, Casilla 19001, Santiago 19, Chile
  		\and
  		Institute for Astronomy, ETH Zurich, 8093 Zurich, Switzerland
  		\and
		Laboratoire d'Astrophysique de Marseille (LAM),13388 Marseille, France
             }

   \date{Received 12 May 2013 / Accepted 4 June  2013}

  \abstract
  {Over the past five years, radial-velocity and transit techniques have revealed a new population of Earth-like planets with masses of a few Earth masses. Their very close orbit around their host star requires an exquisite inner working angle to be detected in direct imaging and sets a challenge for direct imagers that work in the visible range, like SPHERE / ZIMPOL.}  
   {Among all known exoplanets with less than twenty-five Earth-masses we first predict the best candidate for direct imaging. Our primary objective is then to provide the best instrument setup and observing strategy for detecting such a peculiar object with ZIMPOL. As a second step, we aim at predicting its detectivity.}
   {Using exoplanet properties constrained by radial velocity measurements, polarimetric models and the diffraction propagation code CAOS, we estimate the detection sensitivity of ZIMPOL for such a planet in different observing modes of the instrument. We show how observing strategies can be optimized to yield the best detection performance on a specific target.}
   {In our current knowledge of exoplanetary systems, $\alpha$ Centauri B b is the most promising target with less than twenty-five Earth-masses for ZIMPOL. With a gaseous Rayleigh-scattering atmosphere and favorable inclinations, the planet could be detected in about four hours of observing time, using the four-quadrant phase-mask coronograph in the I band. However, if $\alpha$ Centauri B b should display unfavorable polarimetric and reflective properties similar to that of our Moon, it is around 50 times fainter than the best sensitivity of ZIMPOL.}
   {$\alpha$ Centauri B is a primary target for SPHERE. Dedicated deep observations specifically targeting  the radial velocity-detected planet can lead to a detection if the polarimetric properties of the planet are favorable.}
% However planets on larger orbits such as an Earth at 1AU can pass unnoticed among RV data and be detected in polarimetry with ZIMPOL.

   \maketitle
%
%________________________________________________________________

\section{Introduction}

Imaging planets is a very attractive goal to improve our understanding of planetary systems. So far, it has only been achieved in the near-infrared\footnote{Except for Fomalhaut b detected by \citet{Kalas2008} with HST/ACS and confirmed by \citet{Galicher2012} and \citet{Currie2012}, but this is a controversial case because the nature of the object has yet to be revealed.} by detecting the thermal emission of young (1-100 Myr) and massive Jupiter-size planets at large distances from their host stars (5-100 AU). Imaging planets in visible reflected light is also very valuable. However, while the flux reflected by the planet is highest at a very small orbit, the stellar halo is stronger than that of the planet at such a short separation. Moreover, the adaptive optics (hereafter AO) correction is not favorable at visible wavelengths. The contrast required is around $4 \times 10^{-10}$ for an earth at 1 AU from its host star, while the angular separation is only  0.1\arcsec for a star at 10pc.

However, to help detection, a specific property of scattered light can be used: polarization. Polarimetric differential imaging (hereafter PDI) is already widely used to enhance the contrast between a star and circumstellar material, e.g., to reveal protoplanetary disks. Currently, two 8-meter class telescopes provide subarcsec-resolved imaging with a  dual-beam polarimeter: Subaru/HiCIAO and VLT/NaCo. The latter revealed polarized circumstellar emission down to 18 mag/arcsec$^2$ at $1.5\arcsec$ on HD169142 \citep{Quanz2013}. A dedicated instrument for exoplanet search in the visible light will now be installed at the VLT as part of the SPHERE instrument \citep{Beuzit2008}: ZIMPOL, the Zurich IMaging POLarimeter \citep{Schmid2006}. It uses the SPHERE AO system and coronographic masks. ZIMPOL has demonstrated polarimetric sensitivities of $10^{-5}$ locally with an absolute polarimetric accuracy of $10^{-3}$. Fast polarimetric modulation is performed using a ferroelectric liquid crystal to swap two orthogonal linear polarization states at 1 kHz. A polarization beamsplitter converts this modulation into an intensity modulation, which is then demodulated in real-time by a special masked charge-shifting CCD detector. The same CCD pixels are used for the detection of both polarization states to minimize differential effects. Since the modulation period is shorter than the seeing variation timescale, speckle noise is strongly reduced in the polarization image. 

\begin{figure}
   \centering
   \includegraphics[width=\columnwidth]{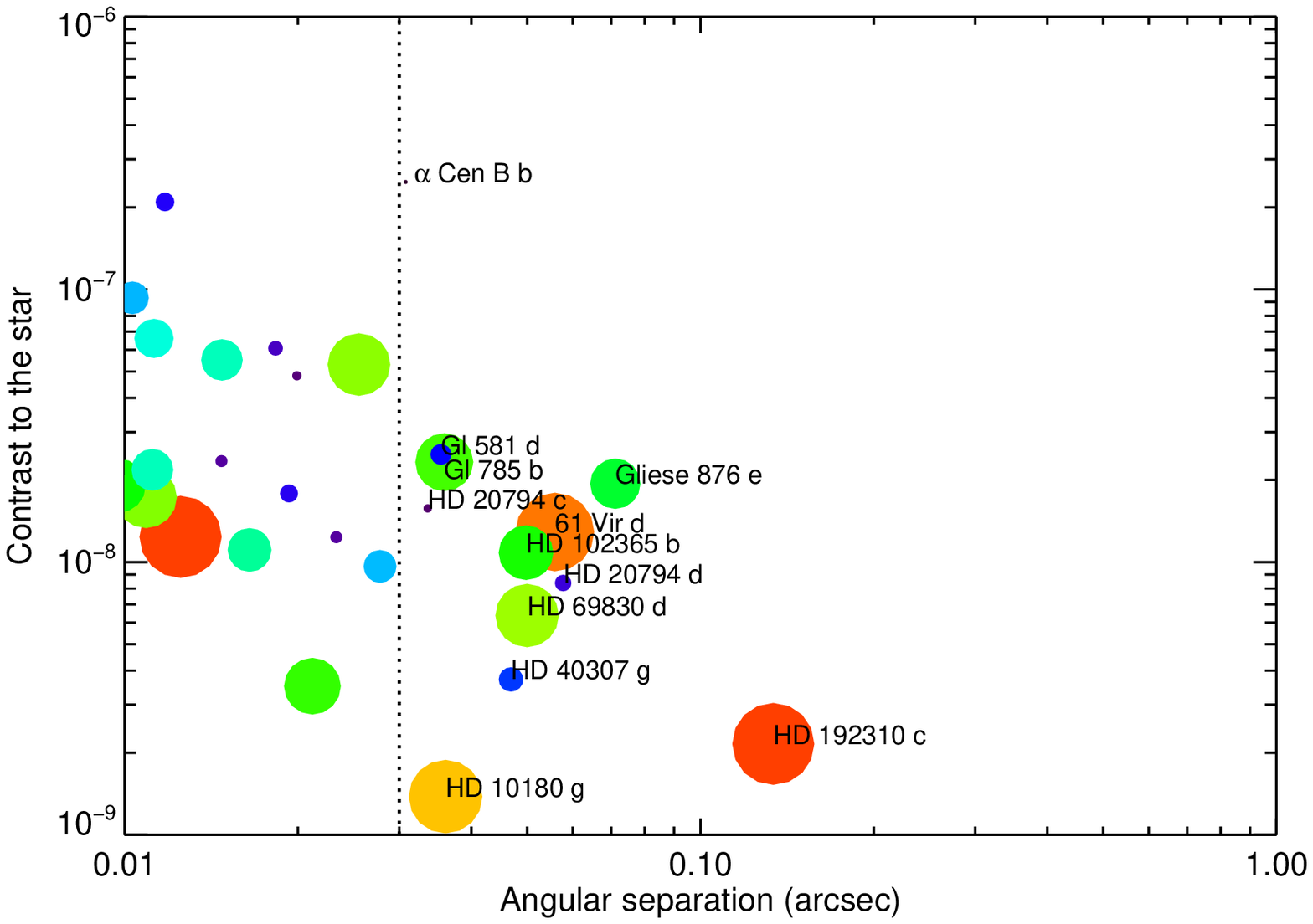}
   \caption{Preliminary intensity contrast of known exoplanets of less than 25 Earth masses confirmed before May 2013 (from exoplanets.eu). The size and color (from blue to red) of the dots are proportional to the planet mass.}
    \label{Fig_intensity_contrast_exoplanets}%
    \end{figure}

The large majority of low-mass exoplanets ($M_{pl} \leq 25 M_{Earth}$) detected in transit or radial velocity (hereafter RV) have a projected angular separation at quadrature smaller than the ZIMPOL inner working angle, however, $2 \lambda / d $ at $ \unit{600}{\nano\metre}$ or $\unit{0.03}{\arcsecond}$. Those with a preliminary intensity contrast higher than one part per billion ($10^{-9}$) and a projected separation larger than $\unit{0.03}{\arcsecond}$ are named in Figure \ref{Fig_intensity_contrast_exoplanets} and constitute our sample selection. The preliminary intensity contrast is given by $f \cdot \left( R_{pl}/{a} \right)^2$ assuming the same reflectance $f=0.2$ for all planets (corresponding to the Rayleigh-scattering atmosphere described below and a scattering angle $\alpha=87^\circ$). The radius $R_{pl}$ is computed from the RV mass $M_{pl} \sin{i} $ assuming an Earth bulk density. The vertical dotted line shows the ZIMPOL inner working angle. Of the 11 targets, $\alpha$ Cen B b is by far the most promising with its intensity contrast of more than $2 \cdot 10^{-7} $. It has a semi-major axis $a=0.04$ AU \citep{Dumusque2012}. Because its parent star is the second-closest star after Proxima Centauri at a distance of 1.34 pc, the projected separation is enhanced but remains small: $\unit{0.03}{\arcsecond}$ at quadrature.

Polarimetric differential imaging is complementary to RV techniques, which have a projection ambiguity because the system inclination $i$ is unknown and only the projected mass $M \times \sin i$ can be determined. This degeneracy can be broken with multi-epoch direct images. 
$\alpha$ Cen B b could then become the first exoplanet to be unambiguously detected both in RV and direct imaging. The only other planet directly imaged with tight RV constraints is $\beta$ Pic b \citep{Lagrange2012_RV}, for which we can constrain its true mass to the range $9-12M_{Jup}$ independently from brightness-mass relations. Moreover, polarimetry gives additional constraints on the planet's atmospheric and surface properties even if those models are highly degenerated.

\section{Expected polarimetric signatures of $\alpha$ Cen B b}
\label{Sec_polar_signature}

The linear polarization is commonly described using the Stokes parameters Q and U. Each of them can be obtained with the difference in intensity between beams with opposite linear polarization ($I_{0^\circ}$ and $I_{90^\circ}$ for Stokes U, $I_{45^\circ}$ and $I_{-45^\circ}$ for Stokes V). 
From RV data,  the planet orbital phase angle can be predicted, but not the system inclination $i$, therefore the planet position angle (PA) on sky is unknown and the direction of linear polarization of the planet cannot be predicted. Consequently, parameters Q and U have to be measured.

For a given observation at time $t$, we call $\phi$ the orbital phase angle of the planet ($\phi=0^\circ$ at inferior conjunction and $\phi=90^\circ$ at quadrature). The remaining unknown parameters upon which the Stokes Q and U parameters depend Stokes Q and U are the following:
\begin{itemize}
\item The system inclination $i$. This will entirely define the scattering angle $\alpha$ of the planet, given by  $\alpha(\phi,i) = \arccos(\cos{\phi} \sin{i} )$.  We assumed that the orbit was fully circularized given the low value of $a$. Four discrete inclinations $10^\circ$, $30^\circ$, $60^\circ$, and $90^\circ$ are used in our simulations, corresponding to a planet true mass of $6.6$, $2.2$, $1.3$, and $1.1$ Earth masses.
\item The disk integrated reflectance $f(\alpha,\lambda)$ and polarization fraction $p(\alpha,\lambda)$ as a function of $\alpha$. The product $f \times p$ is called hereafter polarized reflectance.
\item The planet radius $R_{pl}$ estimated from the mass-radius relation derived for terrestrial planets by \citet{Sotin2007}, with $R_{pl}$ proportional to ${M_{pl}}^{0.274}$. It ranges between $1.7 R_{Earth}$ for a  $10^\circ$ inclination and  $1.0 R_{Earth}$ for an edge-on system. 
\end{itemize}

\begin{figure}
   \centering
   \includegraphics[width=8cm]{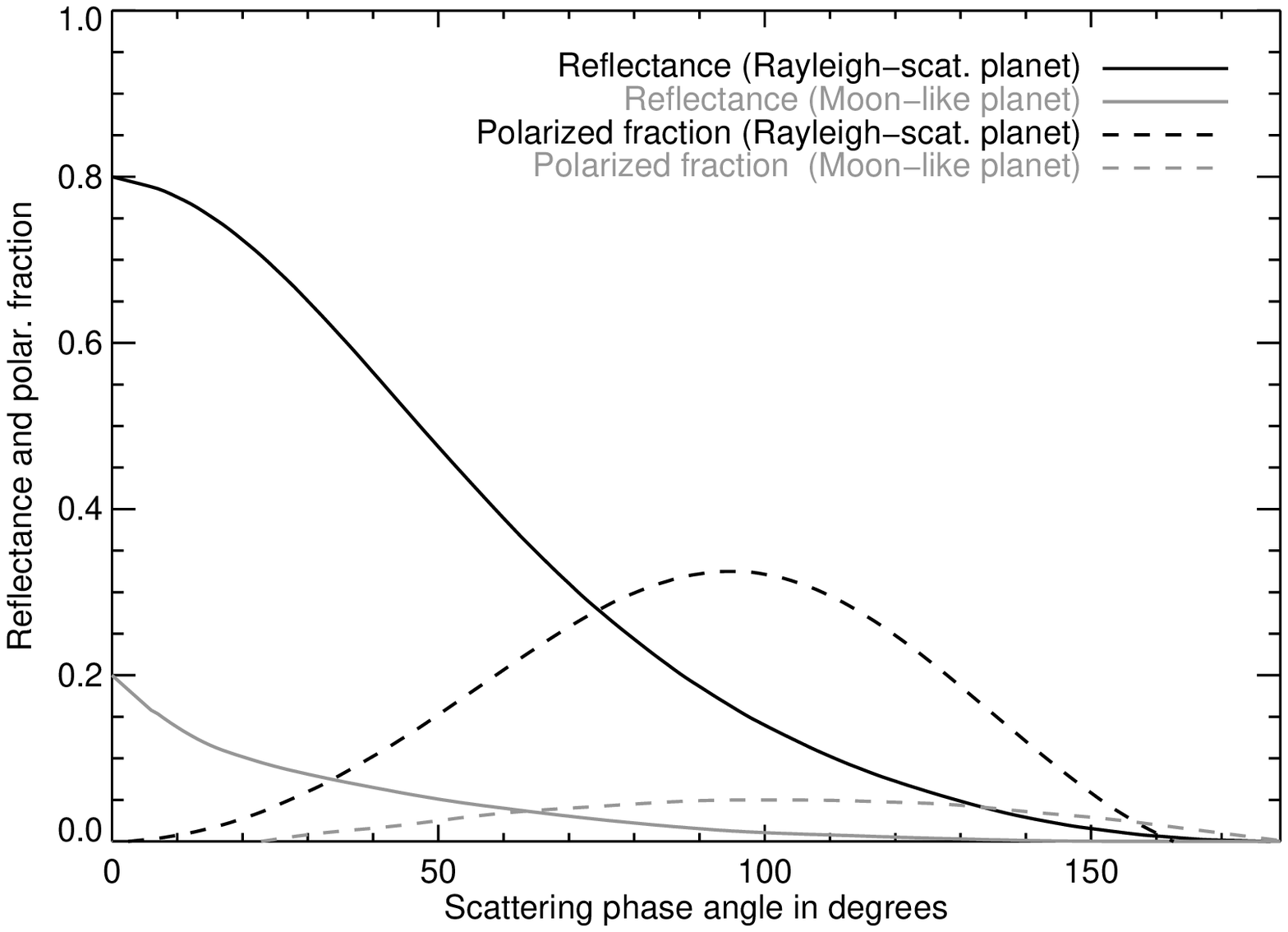}
   \includegraphics[width=8cm]{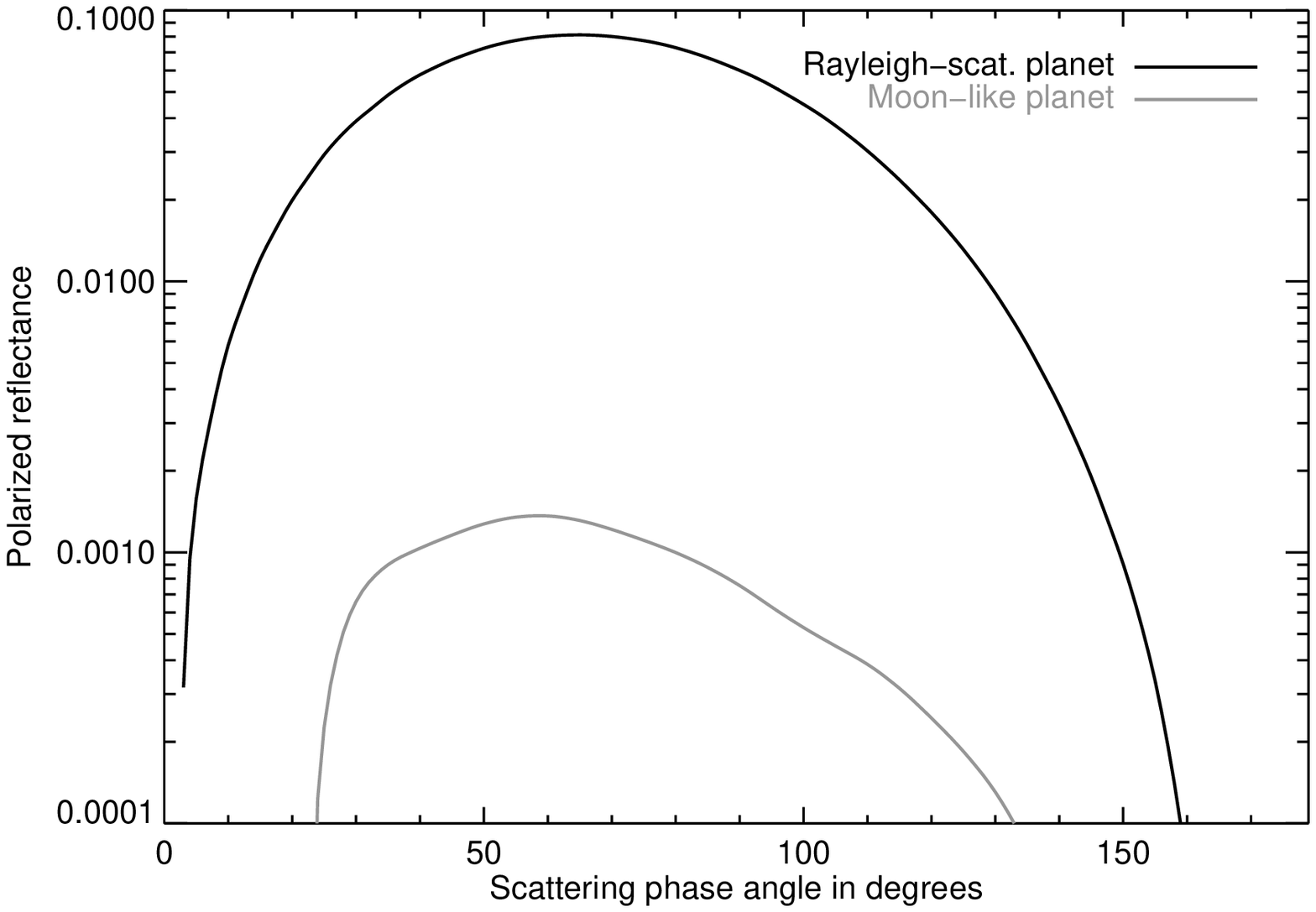}
   \caption{Reflectance $f$ and polarization fraction $p$ (top) and polarized reflectance $f \cdot p$ (bottom)  of the two planet models. For the Moon-like planet, properties are given for the I band.}
   \label{Fig_reflectance_vs_scta_angle}%
    \end{figure}

All in all, the polarimetric contrast is $ p(\phi,i,\lambda) f(\phi,i,\lambda) \left( \frac{R_{pl}}{a} \right)^2 $. We investigate two polarization models for the planet :
\begin{description}
\item[ \textbf{Moon-like planet.}]
This model corresponds to a rocky planet with polarimetric properties like the Moon. Given the small mass and orbital distance of $\alpha$ Cen B b, a tiny atmosphere or even no atmosphere at all is a realistic assumption. The  light is reflected from the solid surface and Moon- or Mercury-like properties are therefore plausible. From a detection point of view this would represent a worst case scenario. We used reflectance and polarization fraction measurements of the Moon derived from \citet{Coyne1970} and \citet{Kieffer2005} for this scenario. They are displayed in Fig. \ref{Fig_reflectance_vs_scta_angle}. The polarized reflectance reaches a maximum of $0.13\%$ for a scattering angle of $60^\circ$ and there is little wavelength dependence in the 600-900nm range. 
 \item[ \textbf{Planet with a Rayleigh-scattering atmosphere.}] 
In more favorable conditions, the planet is assumed to have a rocky core and to retain a Rayleigh-scattering atmosphere that reflects and polarizes much more incident starlight. We used the polarization model presented in \citet{Buenzli2009} for that purpose. It assumes a multiple-scattering atmosphere above a Lambertian surface. It is described by three parameters: the surface albedo $A_s$, the atmosphere total optical depth $\tau$, and the single scattering albedo $\omega$. We chose the most favorable parameter set corresponding to a deep ($\tau= 30$) conservative ($\omega = 1$) Rayleigh-scattering atmosphere above a perfectly reflecting Lambert surface ($A_s=1$). The polarized reflectance peaks at $8.1\%$ for a scattering angle of $63^\circ$. Likely scenarii for such a deep atmosphere would be a planetary surface hot enough to sublimate, explosive volcanism, and/or a slowly evaporating atmosphere.
\end{description}

We did not consider the case of an ocean planet as initially introduced  by \cite{Kuchner2003} and \citet{Leger2004}. Although the polarimetric signature of ocean planets is peculiar because the presence of the specular reflection on the planet's liquid surface in addition to Rayleigh scattering in the atmosphere \citep{Mccullough2006}, runaway greenhouse excludes the possibility that $\alpha$ Cen B b could sustain liquid water on its surface. A runaway greenhouse is indeed possible for an ocean planet at the orbit of Venus \citep{Kopparapu2013}, and $\alpha$ Cen B b receives an insulation as strong as 193 times that received by Venus.

As ZIMPOL is able to perform polarimetry at different visible wavelengths from $\unit{515}{\nano\meter}$ to $\unit{900}{\nano\meter}$, we investigated how chromatic the integrated reflectance and polarization fraction are. $\alpha$ Cen B is a late-type star (spectral type K1V), therefore more scattered flux is expected at longer wavelengths. For our two models, reflectance is also higher at longer wavelengths but the polarization fraction is smaller so that the polarized reflectance is very achromatic over the spectral range and should not be considered a critical item when selecting the wavelength band from an astrophysical point of view. 

For each of the two models, we computed the expected polarized contrast of $\alpha$ Cen B b for different system inclinations $i$ and at different orbital phase angles $\phi$ corresponding to different angular separations of the planet. The value displayed in Table \ref{table_contrast} corresponds to the orbital phase angle that leads to the highest contrast under the constraint that the projected angular separation is greater than $\unit{0.03}{\arcsec}$. As an exercise, we repeated this task for the other ten targets presented in Fig. \ref{Fig_intensity_contrast_exoplanets}, keeping only the favourable Rayleigh-scattering model. This confirms the preliminary result from Fig. \ref{Fig_intensity_contrast_exoplanets}: among low-mass planets ($M_{pl} \leq 25 M_{Earth}$), $\alpha$ Cen B b is an order of magnitude brighter in polarized reflected light than any other potential target for ZIMPOL. The expected polarimetric contrast varies between 1 and 223 ppb, depending on the assumptions.

\definecolor{Gray}{gray}{0.9}
\begin{table}
\caption{Expected contrast in polarized light, expressed in parts per billion ($10^{-9}$) assuming the Rayleigh-scattering atmospheric model, except for $\alpha$ Cen B b where both models are displayed. The contrasts is shown for 4 different system inclinations, except in 2 cases where the inclination is already constrained.}             % title of Table
\label{table_contrast}      % is used to refer this table in the text
\centering                          % used for centering table
\begin{tabular}{ m{3.2cm} | c |  c |  c |  c}        % centered columns (4 columns)
Planet & $i=10^\circ$ & $i=30^\circ$ & $i=60^\circ$& $i=90^\circ$   \\    % table heading 
\hline                        % inserts single horizontal line
\rowcolor{Gray}
 $\alpha$ Cen B b (Rayleigh) &  223 & 119 & 94 & 89 \\        % inserting body of the table
%\rowcolor{Gray}
%$\alpha$ Cen B b (Ocean) & 168  & 93 & 69 & 64 \\
 Gl 581 d & 19  & 12 & 8.8 & 8.1 \\
 Gl 785 b & 16  & 9.8 & 7.3 & 6.7 \\
 HD 20794 c & 13  & 8.4 & 6.2 & 5.7 \\
 Gliese 876 e & \multicolumn{4}{c}{ 6.2 ($i=59.5^\circ$)} \\ 
61 Vir d & 8.4  & 5.3 & 3.9 & 3.6 \\
HD 102365 b & 7.4  & 4.6 & 3.4 & 3.1 \\
HD 20794 c & 6.6  & 4.1 & 3.1 & 2.8 \\
HD 69830 d & \multicolumn{4}{c}{ 3.9 ($i=13^\circ$)} \\ 
\rowcolor{Gray}
$\alpha$ Cen B b (Moon-like) & 3.1  & 1.7 & 1.3 & 1.2 \\
HD 40307 g & 2.8  & 1.7 & 1.3 & 1.2 \\
HD 192310 c & 1.4  & 0.88 & 0.65 & 0.59 \\
\end{tabular}
\end{table}
 
 \section{Observing strategies and instrumental setup}

In a second step, the expected planet polarization signature has to be compared with the ZIMPOL detectivity. This is made by adding the planets polarization signal in a simulated point-spread function (hereafter PSF) produced by the official SPHERE / ZIMPOL simulator \citep{Thalmann2008}. It uses the diffraction code CAOS \citep{Carbillet2008}.  

\subsection{Observing strategies}

To minimize telescope time, we considered an observing scenario where the target is repeatedly observed at the most favorable planetary orbit phase angles with respect to the ZIMPOL sensitivity, namely close to quadrature. The planet position is the same at each visit and the frames can be combined in order to enhance the planet's signal-to-noise ratio (hereafter S/N).
For each visit, the baseline considered is a four-hour observation to measure both the Stokes Q and U parameters, or two hours on each Stokes parameter.
Given the short period of the planet, this is indeed the longest integration that does not lead to a significant planet smearing due to its orbital motion. At the best orbital phase angle of $80^{\circ}$, the smearing of the planet $\alpha$ Cen B b on the detector during a two-hour observation remains below $0.3$ $\lambda / d$ if the system inclination is above $30^{\circ}$, and it is smaller than 0.4 $\lambda / d$ in all cases. Therefore, the expected dilution of the signal was not taken into account in this simulation.

\subsection{ZIMPOL setup}
ZIMPOL's best polarimetric performances are achieved in fast polarimetry: polarimetric modulations are performed at 1kHz, faster than the turbulence timescale. This eliminates most of the stellar halo and its speckle pattern. A quasi-static pattern remains at a level of $10^{-4}$ with respect to the stellar core  because of the low-level optical polarization and the wavefront error variation induced by the polarimetric swap. This pattern is additionally reduced by polarimetric switching introduced by a $45^\circ$ rotation of the half-wave plate (HWP) where the sign of the polarization in front of the switch is reversed, whereas the sign of the instrumental polarization after the switch remains unchanged. This way the static instrumental effects are eliminated by the data reduction process. The remaining level of residuals can be additionally reduced by averaging images corresponding to different offsets of the derotator. This is called active field-rotation. We assumed twelve derotator offsets, which additionally decreased the noise by a factor 3.5. Appendix  \ref{app_caos} quantifies the noise contributors and describes the data reduction steps. The final contrast value is below $10^{-7}$ at $2\lambda / D$. 

\begin{figure}
   \centering
   \includegraphics[width=10cm]{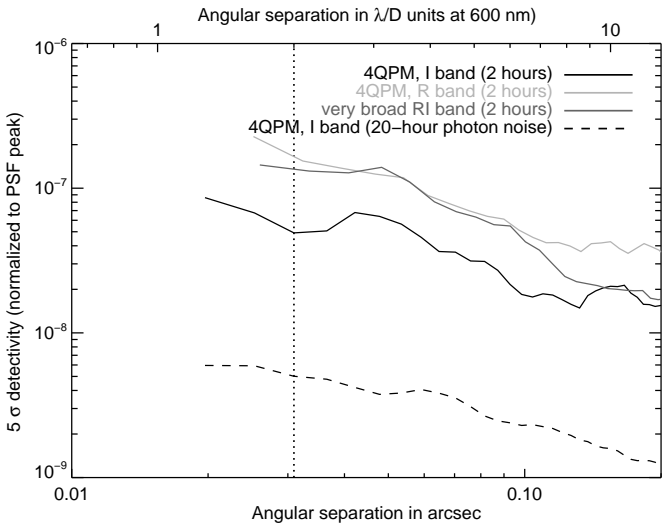}
   \caption{Comparison of the performance of ZIMPOL at close separation in the different bands. Quasi-static, photon, and readout noise are included.}
    \label{Fig_comparison_band}
    \end{figure}

As the expected separation of $\alpha$ Cen B b is $2 \lambda / D$, observations can only be planned with the two 4QPM coronographs or without any coronograph. The Lyot masks do not provide a small enough inner working angle. As the 4QPM are chromatic, two filters are studied here: broad-band I and R, with a central wavelength of $\unit{790}{\nano\meter}$ and $\unit{626}{\nano\meter}$ respectively, and a spectral width of $\unit{150}{\nano\meter}$. The case without coronograph is studied here with the broader filter in ZIMPOL: very broad-band RI covering the full $\unit{590}{\nano\meter}$ to $\unit{880}{\nano\meter}$ spectral range. 
Fig. \ref{Fig_comparison_band} shows the contrast comparison in the different bands. The 4QPM in the I-band gives results more than one order of magnitude better than saturated images or than the 4QPM-R at the separation of $\alpha$ Cen B b. It is more efficient to reject light thanks to a better AO correction (cf Appendix \ref{app_caos}). This advantage more than compensates for the fact that no coronograph allows a broader bandpass.  An additional drawback of not using any coronograph is that saturation is dangerously close to $\alpha$ Cen B b. 

It could be argued that  the presence of the planet at a separation around or slightly below $2 \lambda /D$ could be a problem for detection because this is considered as the inner working angle for the 4QPM coronograph. However, the extinction rate of a companion located at $45^{\circ}$ from a mask transition is relatively constant between $1.7$ and $2 \lambda /D$ \citep{Riaud2001}. 
 
\section{Results}

\begin{figure}
   \centering
   \includegraphics[width=10cm]{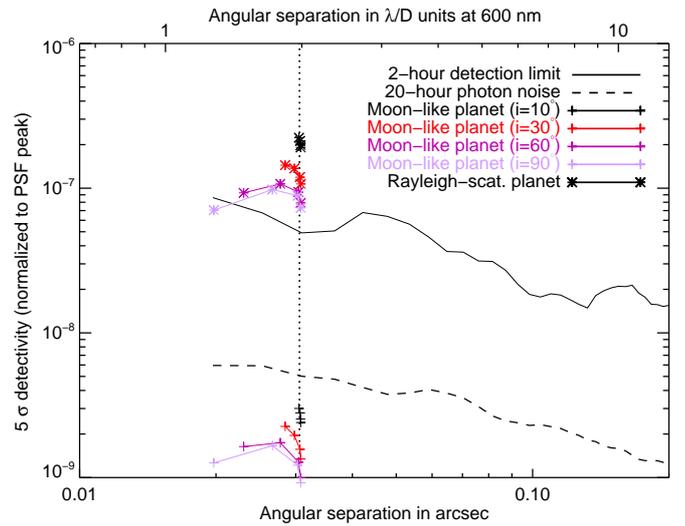}
   \caption{Detection level of the two planet models with the 4QPM coronograph in the I band. The symbols indicate the planet signal for an orbital phase angle $\phi$ of $40^\circ$, $60^\circ$, $80^\circ$, and $90^\circ$ respectively.}
              \label{Fig_4QPM_I_planets}%
    \end{figure}

As shown in Fig. \ref{Fig_4QPM_I_planets}, a rocky planet with a Rayleigh-scattering atmosphere is detected above the $5\sigma$ level whatever the inclination of the system in a total of four hours (two hours for each Stokes parameter Q and U). 
The detection is easier for pole-on systems since the planet's true mass, hence the radius too, is greater. In most cases the detectivity is best for an orbital phase angle $\phi$ between $80^\circ$ and $90^\circ$, therefore the observations should be planned to be conducted in this window. 
For a rocky planet with Moon-like properties, the detection level is about 50 times fainter than the ZIMPOL sensitivity in four hours. For comparison we overplotted the photon noise level for 20 hours of observations. It corresponds to the best possible detectivity level assuming we succeeded in removing all differential aberrations down to that level. A deep understanding of the instrument stability complemented by advanced post-processing techniques will be necessary. 
	 
\section{Conclusions}

We have studied the detectivity of known low-mass exoplanets with the SPHERE / ZIMPOL instrument using two polarization models. Among planets with a mass smaller than $ 25$ Earth masses, $\alpha$ Cen B b is by an order of magnitude the best-known candidate for a direct detection in polarized reflected light. It would be a groundbreaking result unlikely to be surpassed for some time due to the proximity of this exoplanet to Earth. The best setup to observe $\alpha$ Cen B b is to use the fast polarimetric mode of ZIMPOL in the broad-band I filter with the 4QPM coronograph.

A rocky planet with an atmosphere that has ideal Rayleigh-scattering properties can be detected in four hours whatever the inclination but a planet without an atmosphere and with unfavorable scattering properties like the Moon would pass unnoticed. A scenario that could potentially enhance the polarized reflectance is the atmospheric escape that would produce a cometary-tail of ionized gas because the stellar tidal forces extend the Roche limit and the strong radiations heat the planetary surface. The escape of atomic hydrogen was previously revealed for HD 209458 b \citep{Vidal-Madjar2003}, but observational constraints are scarce. These effects may lead to a much stronger polarimetric signature, but a detailed calculation is beyond the scope of this paper.
If the 20-hour photon noise level can be reached by combining several observation epochs and suppressing systematics effects, the $5\sigma$ contrast level is decreased to $5 \cdot 10^{-9} $. This is still above the signal of a Moon-like $\alpha$ Cen B b, but many kinds of planets without ideal Rayleigh-scattering atmosphere become detectable. This also represents the $5\sigma$ contrast level of our ideal Rayleigh-scattering model for a planet with the same mass but a period of 56 days (28 days respectively) if the system is inclined at $10^\circ$ ($90^\circ$ respectively). Such a planet might have passed unnoticed among radial velocity measurements, especially because the star's rotational periods are about 40 days long. Alternatively, with this contrast level, we are now sensitive to much lighter planets whose RV signals are undetectable, so direct imaging will definitely bring a very interesting diagnostic to the planetary system around $\alpha$ Cen B. The results presented here are based on instrument properties as they are known now, but it is clear that more investigations for the best data combination and signal extraction will be pushed forward on the basis of the first on-telescope results.

\bibliography{adsbib} 

%\Online
\begin{appendix} %First online appendix
 
\section{ZIMPOL simulation description}
\label{app_caos} 

A ZIMPOL simulation is conducted using the SPHERE software package for the CAOS  problem-solving environment described in \citet{Carbillet2008}. A comprehensive description of the ZIMPOL simulator is provided in \citet{Thalmann2008}. We briefly summarize the simulation concept and assumptions. The diffraction code of CAOS produces PSFs for the central occulted star and for out-of axis planets, simulating the AO-corrected turbulence with 100 turbulent phase screens and static and differential aberrations. The differential aberrations account for temporal drifts in the common optical path on the timescale of the HWP signal-switching or differences between the two polarimetric channels. The atmospheric and telescope parameters are the same as described in \citet{Thalmann2008}. The assumptions for the static aberrations were updated using data from manufacturing: static instrumental aberrations were decreased from 34.5 to 30nm and AO calibration aberrations were decreased to 5nm. The two simulations of the 4QPM used a total of six wavelengths within the band while the very broad-band simulation used 36 wavelengths. The simulations show a Strehl ratio of about 50\% in the I band with a peak rejection factor slightly above 100. We highlight however that a good AO tip-tilt correction is essential to reach a higher starlight rejection and the 4QPM coronographs display transitions between the quadrants that also degrade the star extinction performance. Additionally, the performance of the atmospheric dispersion corrector is degraded with the broad- and very broad-band filters. 

\begin{figure}
   \centering
   \includegraphics[width=10cm]{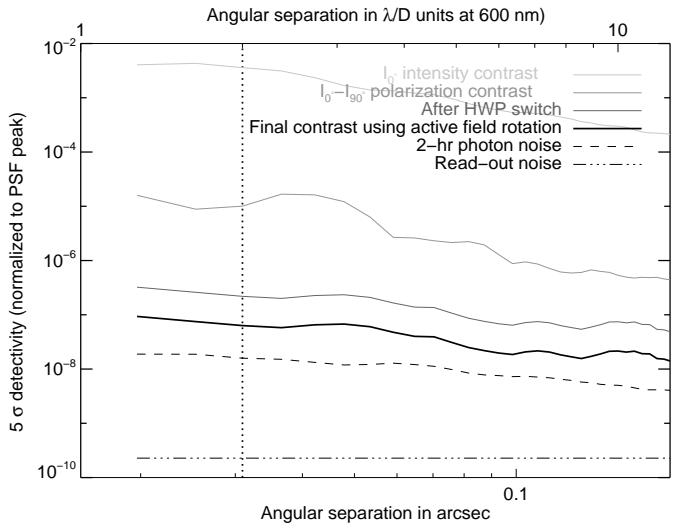}
   \caption{Decomposition of the final polarization contrast curve after the $I_{0^\circ} - I_{90^\circ}$ image subtraction, HWP switch, and active field rotation using 12 derotator positions for a two-hour coronographic observation in the I band. Quasi-static speckles are the dominant noise contributor at short separation.}
              \label{Fig_4QPM_I_noise_decomposition}%
    \end{figure}

In a second step, the PSFs produced by CAOS are combined with the star properties and latest instrumental transmission as measured in the laboratory \citep{Roelfsema2011}. The detector integration time (DIT) was adjusted within its range (0.16 to 10s in the windowed $1\arcsec \times 1\arcsec$ detector mode). With the minimal DIT, the detector is saturated to a level of 3, meaning almost up to the region of interest at $2 \lambda /D$, whereas the typical exposure time with the coronographs is 6s to reach the full dynamic of the detector.
The photon, detector, and polarimetric noise are then added. The detector noise amounts to ten electrons \citep{Schmid2012} for the windowed ($1\arcsec \times 1\arcsec$) readout mode selected, and a polarimetric sensitivity of $10^{-5}$ was used. 
This yields the two final images $I_{0^\circ}$ and $I_{90^\circ}$ with their 2 analogs once the HWP has been switched.  We assumed that this switch was performed every five minutes and the resulting temporal aberrations are the same as described in \citet{Thalmann2008}. The typical contrast level of the four images is shown in Fig. \ref{Fig_4QPM_I_noise_decomposition} (second curve from the top). The following steps are then performed:  
\begin{itemize}
\item Subtraction of the intensity images $I_{0^\circ}$ from $I_{90^\circ}$ (second curve from the top in Fig. \ref{Fig_4QPM_I_noise_decomposition}).
\item Subtraction of the resulting image from its analog after the HWP switch by $45^{^\circ}$ (third curve from the top in Fig. \ref{Fig_4QPM_I_noise_decomposition}).
\item Accounting for contrast improvements due to active field rotation. At a separation of 2$\lambda / D$, there are $2\times 2\pi=12$ resolution elements. Assuming there are independent realizations of a Gaussian process, a pattern with 12 derotator offsets would additionally decrease the remaining noise by a factor $\sqrt{12}=3.5$ (darkest curve in Fig. \ref{Fig_4QPM_I_noise_decomposition}).
\end{itemize}

This simulation mainly relies on two consecutive image differences: the first one between two instantaneous images with two orthogonal polarization directions and a second one between images with different positions of the HWP separated by about five minutes. In practice, this second difference could be achieved by building the image that best matches the current one from a library of PSFs. When the instrument shows a stable behavior over time, this library could contain PSFs corresponding to telescope realizations at a different date or on different stars. Advanced image-processing algorithms could then be used to build the best-matching PSF and reduce the remaining aberrations even more. 

\end{appendix}

\end{document}